Original research

# Modeling the onset of photosynthesis after the Chicxulub asteroid impact

[1]Noel Perez, [1]Rolando Cardenas, [1]Osmel Martin, [2]Reinaldo Rojas

[1] Department of Physics, Universidad Central de Las Villas, Santa Clara, Cuba. Phone 53 42 281109. Fax 53 42 281130. e-mail: noelpd@uclv.edu.cu, rcardenas@uclv.edu.cu, osmel@uclv.edu.cu

[2] Museo Nacional de Historia Natural, La Habana, Cuba. e-mail: rojas@mnhnc.inf.cu

**Abstract** We do a preliminary modelling of the photosynthetic rates of phytoplankton at the very beginning of the Paleogene, just after the impact of the Chicxulub asteroid, which decisively contributed to the last known mass extinction of the Phanerozoic eon. We assume the worst possible scenario from the photobiological point of view: an already clear atmosphere with no ozone, as the timescale for soot and dust settling (years) is smaller than that of the full ozone regeneration (decades). Even in these conditions we show that most phytoplankton species would have had reasonable potential for photosynthesis in all the three main optical ocean water types. This modelling could help explain why the recovery of phytoplankton was relatively rapid after the huge environmental stress of that asteroid impact. In a more general scope, it also reminds us of the great resilience of the unicellular biosphere against huge environmental perturbations.

**I Introduction**

Asteroid impacts are a serious threat for life on Earth, and very likely also for biospheres in exoplanets. Specifically, the Chicxulub asteroid impact is widely accepted as the main contributor to the mass extinction in the Cretaceous-Paleogene boundary, which claimed the life of dinosaurs and in general of roughly 50 % of (easily observed) living genera. There were several environmental stresses, and the most accepted scenario immediately after the impact is the ''cold and darkness'' one: aerosols, soot and dust in the atmosphere totally covered sunlight at least during half a year, with the consequent collapse of photosynthesis and a global deforestation. However, it seems that soon after the dust settled and sunlight made it through the atmosphere, a significant recovery of phytoplankton took place. Recent work gives some clues on how this could happen, showing that some species of phytoplankton can grow after decades of dormancy (Ribeiro et al. 2010).

The ozone layer was totally destroyed, due to the release of great quantities of chlorine and bromine from evaporation of both the asteroid and target rocks. After the atmospheric dust settled and photosynthetically active radiation (PAR) reached the ground, the photobiological regime at planetary surface would still be crude, now due to the increased solar ultraviolet radiation (UVR), more or less similar to that of the Early Archean. In this work we apply a mathematical model of photosynthesis to assess the efficiency of phytoplankton photosynthesis in those conditions.

**II Materials and methods**

The atmospheric model used is basically an Archean ozone-less one, giving the worst possible scenario at sea surface for the beginning of Paleogene (from the photobiological point of view). Thus, the solar spectral irradiances at sea level were similar to those used in (Cockell 2000), for solar zenith angles (sza) of 0 and 60 degrees. However, there is strong evidence that the Archean ocean was very clear (Cockell 2000), while there are uncertainties concerning the ocean optical quality one year after the Chicxulub impact (estimated timescale for atmospheric dust to settle). It is likely that it was a turbid ocean, but in this work we use a general optical ocean water classification, allowing us to practically consider the full range of possibilities (Jerlov 1976; Shifrin 1988).

The spectral irradiances $E(\lambda,z)$ at depth $z$ in the water column were calculated using the Lambert Beer's law of Optics:

$$E(\lambda,z) = E(\lambda,0^-)\exp[K(\lambda)z] \quad (1),$$

where $K(\lambda)$ are the attenuation coefficients (defining the kind of optical water type) and $E(\lambda,0^-)$ are spectral irradiances just below sea surface. They are calculated substracting the reflected irradiances from the incident ones $E(\lambda,0^+)$:

$$E(\lambda,0^-) = [1-R]E(\lambda,0^+) \quad (2),$$

In the above expression $R$ is the reflection coefficient, estimated with the help of Fresnel formulae applied to the interface air-water.

Total irradiances $E_{PAR}(z)$ at depth $z$, for the case of photosynthetically active radiation (PAR), are calculated by:

$$E_{PAR}(z) = \sum_{\lambda_I}^{\lambda_F} E(\lambda, z)\Delta\lambda \quad (3),$$

with $\lambda_I = 400nm$ and $\lambda_F = 700nm$ being the extremes of the PAR band and $\Delta\lambda$ is the width of the intervals between the wavelengths for which $K(\lambda)$ were actually measured. The oceanologist N. Jerlov carefully measured $K(\lambda)$ at intervals $\Delta\lambda = 25nm$ in many water bodies, in order to do his optical classification of ocean waters (Jerlov 1976; Shifrin 1988). However, based on this, some of us made linnear interpolation to get $K(\lambda)$ nanometer per nanometer, thus in our case $\Delta\lambda = 1nm$ (Peñate et al 2010). For the case of the (inhibitory) ultraviolet band, spectral irradiances are convolved with a biological action spectrum $\varepsilon(\lambda)$, which weights the biological effect of each wavelength of the ultraviolet band:

$$E^*_{UV}(z) = \sum_{\lambda_I}^{\lambda_F} \varepsilon(\lambda) E(\lambda, z)\Delta\lambda \quad (4),$$

The asterisk in $E^*_{UV}(z)$ means that it is a biologically effective irradiance, as the physical one was convolved (weighted) with a biological action spectrum.

Finally, to account for the photosynthesis rates $P$ (normalised to the maximum rates $P_S$), we used the so called $E$ photosynthesis model for phytoplankton, which assumes good repair capabilities to UVR damage (Fritz et al. 2008):

$$\frac{P}{P_S}(z) = \frac{1 - e^{-E_{PAR}(z)/E_S}}{1 + E^*_{uv}(z)} \quad (5),$$

where $E_S$ is a parameter indicating the efficiency of the species in the use of PAR, inversely proportional to the quantum yield of photosynthesis: the smaller $E_S$, the more efficient the species is. We sampled $E_S$ in a very wide range, spanning from 5 W/m² up to 150 W/m². Certainly, most (if not all) current species fall within this range.

Summing up, the system of equations (1-5), rather than being solved, is computed: having the solar spectrum $E(\lambda, 0^+)$ just above sea surface, we obtain the spectrum just below substracting the reflected light, (eq. (2)). Then the light field $E(\lambda, z)$ down the water column is obtained through the well known Lambert-Beer's law (eq.(1)). Total visible and ultraviolet irradiances $E(z)$ are then computed as a sum of spectral irradiances at depth $z$ (eqs.(3-4)) and finally they are used to calculate photosynthesis rates (eq. (5)).

## III Results and discussion

The figures below illustrate the relative photosynthesis rates in the first 200 meters of the water column.

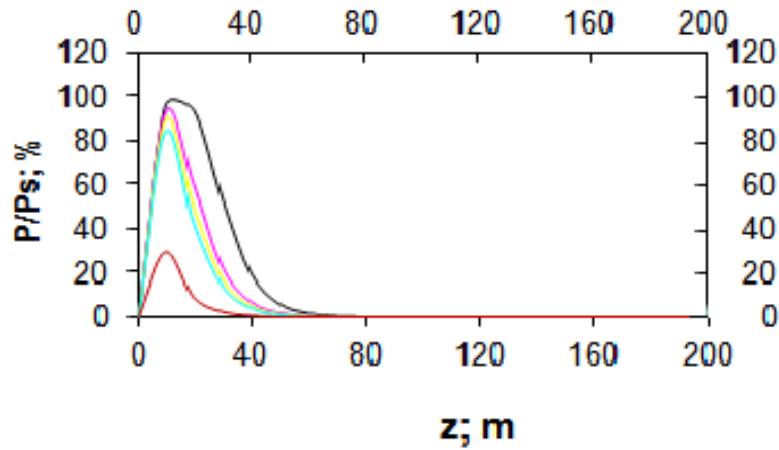

**Fig.** 1 Relative photosynthesis rates for water type III (turbid), with solar zenith angle 0 degree. The values of the parameter $E_S$ are 5 W/m$^2$ (black), 15 W/m$^2$ (magenta), 20 W/m$^2$ (yellow), 25 W/m$^2$ (green) and 150 W/m$^2$ (brown).

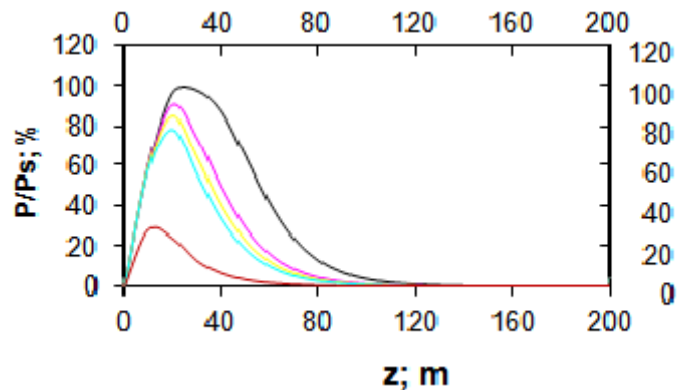

**Fig.** 2 Relative photosynthesis rates for water type II (intermediate), with solar zenith angle 0 degree. The values of the parameter $E_S$ are 5 W/m$^2$ (black), 15 W/m$^2$ (magenta), 20 W/m$^2$ (yellow), 25 W/m$^2$ (green) and 150 W/m$^2$ (brown).

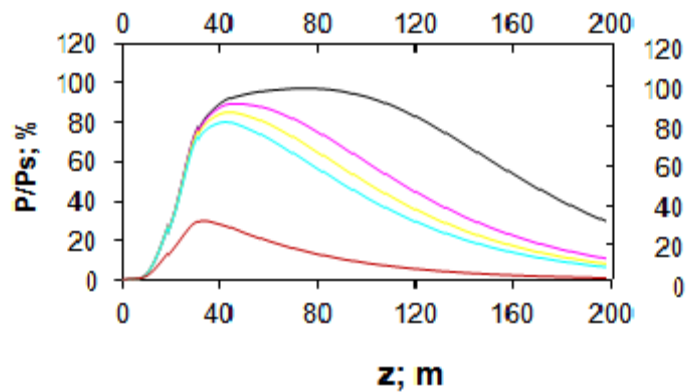

**Fig. 3** Relative photosynthesis rates for water type I (clear), with solar zenith angle 0 degrees. The values of the parameter $E_S$ are 5 W/m² (black), 15 W/m² (magenta), 20 W/m² (yellow), 25 W/m² (green) and 150 W/m² (brown).

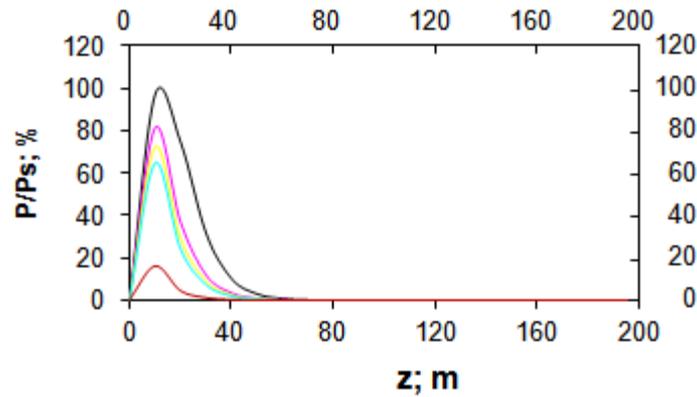

**Fig. 4** Relative photosynthesis rates for water type III (turbid), with solar zenith angle 60 degrees. The values of the parameter $E_S$ are 5 W/m² (black), 15 W/m² (magenta), 20 W/m² (yellow), 25 W/m² (green) and 150 W/m² (brown).

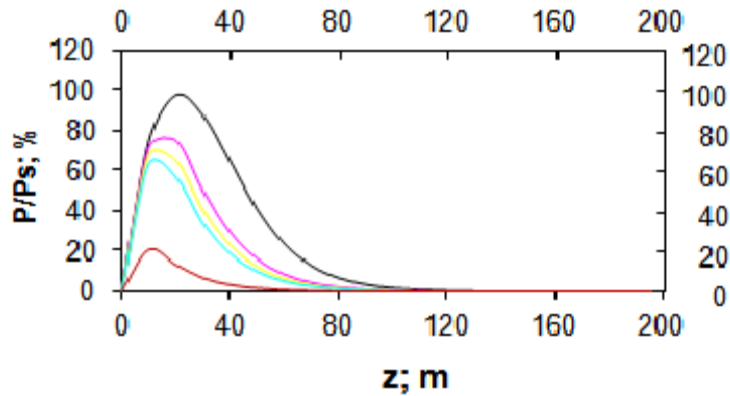

**Fig. 5** Relative photosynthesis rates for water type II (intermediate), with solar zenith angle 60 degrees. The values of the parameter $E_S$ are 5 W/m² (black), 15 W/m² (magenta), 20 W/m² (yellow), 25 W/m² (green) and 150 W/m² (brown).

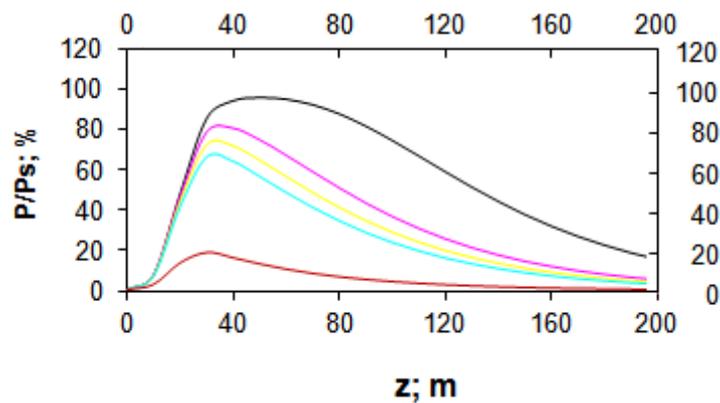

**Fig. 6** Relative photosynthesis rates for water type I (clear), with solar zenith angle 60 degrees. The values of the parameter $E_S$ are 5 W/m² (black), 15 W/m² (magenta), 20 W/m² (yellow), 25 W/m² (green) and 150 W/m² (brown).

In all cases the maximum potential for photosynthesis is up 100% for highly efficient species ($E_S \sim 5$ W/m²), 70-90% for intermediate ones ($E_S \sim 15$-25 W/m²) and 10-30% for the low efficient ($E_S \sim 150$ W/m²). The depth at which the maximum is achieved depends on the balance of inhibitory ultraviolet radiation (preferentially attenuated in the first tens of meters of the water column) and the photosynthetically active radiation, which reaches deeper in the ocean.

However, above plots do not give an accurate enough idea on the viability of photosynthesis, as circulation in the upper mixed layer of the ocean exposes phytoplankton to varying levels of irradiation. To account for this, we consider a simple pattern of circulation in the upper ocean: circular vertical Langmuir currents with constant speed. Thus, we split a heuristic 40 meters deep mixed layer (a Langmuir cell) in smaller layers with 2 meters thickness and use:

$$\left\langle \frac{P}{P_S} \right\rangle = \frac{\sum_{i=1}^{20} \left\langle \frac{P}{P_S} \right\rangle_i}{20} \qquad (6),$$

where *i* represents the *i-th* layer. Results are shown in the tables below.

|                | $E_S$=5 W/m² | $E_S$=15 W/m² | $E_S$=20 W/m² | $E_S$=25 W/m² | $E_S$=150 W/m² |
|----------------|------|------|------|------|------|
| **Water type I**   | 47,1 | 43,4 | 40,2 | 37,1 | 11,1 |
| **Water type II**  | 71,3 | 51,1 | 44,5 | 39,4 | 10,0 |
| **Water type III** | 53,3 | 34,0 | 29,2 | 25,6 | 6,4  |

**Table 1** Average photosynthesis rates in a 40 m depth Langmuir cell (sza=60 degrees)

|                | $E_S$=5 W/m² | $E_S$=15 W/m² | $E_S$=20 W/m² | $E_S$=25 W/m² | $E_S$=150 W/m² |
|----------------|------|------|------|------|------|
| **Water type I**   | 38,7 | 38,4 | 37,7 | 36,7 | 15,5 |
| **Water type II**  | 73,2 | 61,5 | 56,2 | 51,6 | 16,3 |
| **Water type III** | 62,7 | 44,2 | 39,1 | 35,2 | 10,5 |

**Table 2** Average photosynthesis rates in a 40 m depth Langmuir cell (sza=0 degree)

We see that phytoplankton with intermediate and highly efficient photosynthetic apparatuses (given by $E_S$) would have had reasonably good chances to thrive in an ocean under an ozoneless atmosphere, just one year after the Chicxulub impact.

**IV Conclusions**

It has been shown that after the atmosphere cleared, even in the worst scenario of absolutely no atmospheric ozone, phytoplankton would have good chances to recover due to the protective action of ocean water, no matter its type. In turbid (eutrophic, type

III waters) recovery would be necessarily in the upper 50 meters of the ocean, as deeper there would not be enough photosynthetically active radiation. In clear (oligotrophic, type I waters) in principle all the photic zone (assumed to be 200 meters deep) would be adequate. On the other hand, photosynthesis in intermediate waters (mesotrophic, type II) would be possible roughly in the first 100 meters of the water column.

This work only addresses the photobiological side of the huge perturbation that the Chicxulub impact represented. Indeed, there were several other environmental stresses which could modify the above presented picture. One of them is the potential chemical poisoning of waters due to the global fires. However, it is a fact that phytoplankton quickly recovered, and here we show that from the purely photobiological point of view, most species would have had reasonable potential for photosynthesis just after the atmosphere cleared, for all the three main optical types of ocean water, and even in the worst case scenario of absolutely no atmospheric ozone.

This modelling could help explain why the recovery of phytoplankton was relatively rapid after the huge environmental stress of that asteroid impact. In a more general setting, it also reminds us of the great resilience of the unicellular biosphere against huge environmental perturbations.